\documentclass[twocolumn,final]{article}
\usepackage{epsfig}
\usepackage{harvard}
\usepackage{graphicx}
\usepackage{longtable}

\title{Determination of optimal effective interactions between amino acids 
in globular proteins.\\
{\em \small (Optimal effective interactions of amino acids)}}
\author{Giovanni Settanni$^1$,
Cristian Micheletti$^1$, Jayanth Banavar$^2$, Amos Maritan$^1$\\
\\(1)International School for Advanced Studies (SISSA) and
INFM,\\
Via Beirut 2, 34014 Trieste, Italy;\\
The Abdus Salam Centre for Theoretical Physics - Trieste, Italy;\\
(2)Department of Physics and Center for Materials Physics, 104 Davey
Laboratory,\\ The Pennsylvania State University, University Park, Pennsylvania
16802
}

\begin{document} 
\maketitle 

\begin{abstract} 

An optimization technique is used to determine the pairwise interactions
between  amino acids in globular proteins. A numerical strategy is applied to a
set of proteins for maximizing the native fold stability  with respect to
alternative structures obtained by gapless threading. The extracted parameters
are shown to be very reliable for identifying the native states of proteins
(unrelated  to those in the training set) among thousands of conformations.
The only poor performers are proteins with heme groups and/or poor
compactness  whose complexity cannot be captured by standard pairwise energy
functionals.

\end{abstract}

\newcommand {\eq}[1]{\begin{equation} #1 \end{equation}}
\newcommand {\eqa}[1]{\begin{eqnarray} #1 \end{eqnarray}}

\section{Introduction} 
A knowledge of the interaction potentials between
amino acids is of crucial importance both for predicting the three-dimensional
structure of a protein's native state   and for designing novel proteins,
folding on a desired target conformation
\cite{BB94,BL91,BL93,CS92,GK92,GL92,SO94,HS95,JT92,OS93,BE94,DA94,KS94,LE76,LE83,SK93,SU93,WU94,DK96,SR95,MS98a,SMM98,CRE,BRTO,AN73}.
This builds on the assumption that the interactions between amino
acids (and the solvent) are principally responsible for driving the
folding of a protein to its native state. This is supported by
considerable experimental evidence that the native states of many
globular proteins correspond to free energy minima
\cite{AN73,WO95,CRE,BRTO}. 

On a microscopic scale, all-atom potentials are used to carry out
``first principle" molecular dynamics for folding \cite{VGUN}. Due to
the high level of details included in such calculations, the folding
processes of short peptides can be followed only for rather short
time-scales (of the order of 1 $\mu$s as in \citeasnoun{DK98}). While
the impact of such ``ab initio" calculations is destined to grow rapidly, at
present, highly satisfactory results can be obtained by adopting
mesoscopic phenomenological approaches. Within this framework, a
commonly used approach is to avoid a detailed description of an
amino acid but represent it as a sphere or an ellipsoid centered on
the $C_{\alpha}$ or $C_{\beta}$ position
\cite{MC92,SR95,KGS93,SB95,MS98a}.  This coarse-grained procedure
amounts to integrating out the fine degrees of freedom of a peptide
chain and introduces effective interactions between the surviving
degrees of freedom.

One commonly used strategy to extract coarse grained potentials
between pairs of amino acids has been proposed by
\citeasnoun{MJ85}. The method is based on  the quasichemical
approximation and it entails the calculation of pairing frequencies of
amino acids observed in native structures of naturally occurring
proteins. Similar approaches have been reviewed by \citeasnoun{SI95} and
\citeasnoun{WR93}.  \citeasnoun{TD96} have recently tested the
validity of this procedure on exactly solvable lattice models for
proteins. In all the cases they considered the extracted potentials
did not correlate too well with the true potentials, although the two
sets shared a common trend.

A different strategy for extracting potentials was suggested by
\citeasnoun{MC92} and recently an optimized version has been introduced
\cite{VC98,SMB98}. Rigorous tests, similar to the ones in \citeasnoun{TD96},
carried out both for lattice and off-lattice models have shown that the
optimized strategy converges to the exact potentials for increasing chain
length and/or number of proteins in the training set. The method, explained in
detail in section \ref{methods}, uses the following basic ingredients: the
potentials parametrizing a suitably chosen Hamiltonian must be such that the
energy of a protein sequence in its own native state is lower than in any other
alternative conformations that the protein can attain. For each sequence this
yields a set of linear inequalities involving the unknown interaction
potentials. Two key points need to be addressed carefully when applying this
procedure: the parametrization of the Hamiltonian and the generation of
alternative conformations. If the parametrization of the energy is too poor
and/or  there are unphysical conformations among the decoys (i.e. violating
steric contraints), then no consistent solution can be found (unlearnable
problem, for an example see \citeasnoun{VC98} and \citeasnoun{VN99}) . On the
other hand, if the parametrization is reliable and there are no unphysical
decoy conformations, the energy parameter satisfying the inequalities lie
in a convex region of parameter space. While all points inside the cell satisfy
the whole set of inequalities, there is an optimal point, typically equidistant
from the hyperplanes bounding the cell.  The potential parameters corresponding to the
optimal point, ensure that the native states of proteins are maximally stable
with respect to alternative structures. Our strategy aims at pinpointing the
optimal solution, while the original Maiorov and Crippen strategy stopped when
reaching an unspecified sub-optimal point inside the cell. Our approach differs
from the one employed in \cite{MC92} also because of the different interaction matrix:
in our scheme (as in \citeasnoun{MJ85} or \citeasnoun{KGS93}) the interaction
energy of amino acids pairs does not depend on their sequence separation, while
a complementary strategy was followed by Maiorov and Crippen. In the next
Section we introduce the coarse grained model for proteins and give an overview
of the optimal potential extraction technique. The latter is discussed in
detail in Section \ref{methods}. An assessement of the performance of extracted
potentials, and a comparison with previously known interactions, are given in
Section \ref{sec:res}.

\section{Theory}

\subsection{The Model}

We choose not to introduce any subdivision of amino acids in classes
and retain the full repertoire of 20 types. As is
customary, we used a simplified representation of protein structures
and replaced amino acids with a centroid placed at the $C_\beta$
position \cite{SR95}. A fictitious $C_\beta$ was constructed for
glycine and for amino acid entries without it, by using standard
rotamer angles following \citeasnoun{PL96}.

The basic assumption is that the stable structure of a protein is determined by
several factors, that can be ultimately reduced, through an averaging process,
to effective contact interactions between amino acids. Thus, we postulate the
existence of a functional of the contacts between protein residues, which is in
correspondence with the protein energy. The values attained by such a functional
should relate to the degree of stability of the conformations housing the
sequence.

The strength of a contact between two amino acids whose $C_\beta$'s are
at positions $r_ 1$ and $r_ 2$ is defined according to the following
form, which is a smooth approximation to a stepwise contact function with cutoff
at 8.0\AA:

\eq{\Delta(r_ 1 , r_ 2)= \tanh \left( \left(8.0- |r_ 1 - r_ 2|
\right)/2\right) /2 +0.5 \ .}

\noindent The smooth nature of $\Delta(0,r)$ ensures that our results are
not very sensitive to the actual form of $\Delta(0,r)$. For simplicity of notation, in the following, we
will indicate contact maps with the symbol $\Delta$.

Two Hamiltonian forms for the energy of a sequence $S$ on a
structure $\Gamma$ were considered. First we adopted the following
contact energy function:

\eq{\label{nrgns} E(S,\Gamma)=\sum_{i>j+1}^ N \epsilon ({A_ i},{A_ j}) \cdot
\Delta(r_ i, r_ j)\ ,} 

\noindent where the sum is over all pairs of non-consecutive residues, $ N$ is
the protein length and ${A_ i}$ is the  amino acid type (there are altogether 20
types) at $r=r_ i$.
$\epsilon$ is the $20\times 20$ matrix of contact energies. Since $\epsilon$ is
symmetric, there are only $210$ distinct entries in the matrix. We also
considered a second form with 20 additional terms related to the degree of
solvation of amino acid types:

\eqa{\label{nrgs} E(S,\Gamma)=\sum_{i>j+1}^ N \epsilon ({A_ i},{A_ j})
\cdot \Delta(r_ i, r_ j)+ \nonumber \\ +\sum_{i=1}^ N \epsilon(0,{A_
i})\cdot\sum_{{\scriptsize \begin{array}{c} j=1 \\ {j\neq i, i\pm 1}
\end{array}}}^ N \Delta(r_ i, r_ j)\ .}

The very last sum in (\ref{nrgs}) corresponds to the total number of
contacts of the $i$th residue and reflects its degree of burial.
Accordingly, polar amino acids, typically residing at a protein's
surface are expected to have solvation parameters, $\epsilon(0,A)$,
larger than the hydrophobic ones. Expression (\ref{nrgs}) is formally
equivalent to (\ref{nrgns}) in that it can be rearranged to obtain a
unique sum involving just $210$ terms:

\eq{ E(S,\Gamma)=\sum_{i>j+1}^ N \left[ \epsilon ({A_ i},{A_ j}) +
\epsilon (0,{A_ i})+ \epsilon (0,{A_ j})\right] \cdot \Delta(r_ i, r_
j) .}

Nevertheless, using our strategy to extract energy parameters,
expressions (\ref{nrgns}) and (\ref{nrgs}) turn out not to be 
equivalent. In expression (\ref{nrgs}), the coefficients multiplying
$\epsilon(0,A)$ are large with respect to those pertaining to the
general $\epsilon(A,A^\prime)$ entries. The solvation term will
accordingly give a significant contribution to the energy of a
sequence. This feature was shown to be very useful to discriminate the
native state of a protein from decoy structures
\cite{PL96,DM97}. Furthermore, by using (\ref{nrgs}), it is possible to
estimate the solvent-amino acid interaction, a procedure not carried out
by \citeasnoun{MC92}.

The interaction parameters appearing in eq. (\ref{nrgns}) and
(\ref{nrgs}) are not completely independent since the energy scale can
be fixed arbitrarily \footnote{In other potential extraction schemes,
the potentials are shifted to make their average zero. {\em A
priori}\/ this may not be allowed, since the energy shift will
typically affect the average protein solubility \cite{GMp}.}. To remove
this degree of freedom, we choose to set the norm of the vector
describing the potentials to 1,
\begin{equation}
{\sum_{A\le A^{\prime}} \epsilon^2(A,A^{\prime})=1\quad.}
\label{eqn:norm}
\end{equation}
\subsection{Optimal strategy}

The key prescription at the heart of the potential extraction scheme
is that a protein sequence attains the lowest possible energy when
mounted on its correct native state. Hence, assuming that the energy
parametrizations (\ref{nrgns}) and (\ref{nrgs}) are reliable, the
correct potentials will be such that the native state has the lowest
energy when compared to alternative conformations.

The first step of the analysis was to compile a list of non-homologous proteins
representing a variety of folds (see section \ref{methods} for details). For
each protein sequence in this training set, $S_i$ (with known native state
$\Gamma_i$), the alternative structures are obtained by threading on
conformations in the training set of equal or longer length \cite{JT92}. Thus,
for the correct set of potentials:

\eq{ \label{inq}E(S_i,\Gamma_i) < E(S_i,\Gamma_D)\ ,}

\noindent for all the decoy structures, $\Gamma_D$, obtained by
threading. Therefore, for each sequence in the training set, one obtains
an array of inequalities. Due to the finite number of proteins in the
training set, the whole ensemble of inequalities will be satisfied by
more than a single set of potentials. Indeed, there will typically be
a whole region of points in parameter space each corresponding to a
set of potentials consistent with inequalities (\ref{inq}). The
optimal solution is attained by simultaneously maximizing the
stability gap for all proteins in the set. The stability gap is
defined as the smallest energy difference between a protein's native
state and one of the decoy conformations. The optimal stability
requirement implies that the following inequalities should hold
simultaneously for each training protein

\eq{\label{ineq} \frac{E(S_i,\Gamma_{D})-E(S_i,\Gamma_{i})}{f\left(D\left(
\Gamma_{D},
\Gamma_{i}\right)\right)}>c \quad \quad \forall \quad \Gamma_{D}\ ,}

\noindent where $c$ is a positive quantity to be made as large as possible,
the $\Gamma_D$'s belong to the set of decoy conformations and the energy
interactions satisfy to (\ref{eqn:norm}).

The function $f$ in the denominator of (\ref{ineq}) is a function of the
structural distance between $\Gamma_D$ and $\Gamma_i$. This serves
the purpose of making inequalities (\ref{ineq}) more stringent when
mounting $S_i$ on structurally dissimilar conformations. 
\noindent We used three different trial functions for $f$:

\eqa{f_1(x)&=&1 \label{fform}\ ,\\
f_2(x)&=&x\label{sform}\ ,\\
f_3(x)&=&x^2\label{tform}\ .}

\noindent For the distance function $D(\Gamma,\Gamma^\prime)$, appearing in eq.
(\ref{ineq}), we used the Euclidean distance in contact-map space:

\eq{ D\left( \Gamma, \Gamma^\prime \right) \equiv\left[\frac{
\sum_{i>j+1=2}^ N (\Delta(r_ i, r_ j)-\Delta(r^\prime_ i, r^\prime_ j))^
2}{(N-1)(N-2)/2} \right]^{1/2}.}

\noindent $D(\Gamma,\Gamma^\prime)$ can be viewed as a close relative  in terms
of contact maps of the standard distance root mean square deviation (DRMSD) but
related to our definition of the energy functional.

By threading the training sequences on longer structures, we generated the whole
set of inequalities (\ref{ineq}). Each of these identifies a hyperplane in
parameter space dividing space into two semi-infinite regions; one of which is
compatible with the inequality and contains the physical set of parameters
\cite{VC98}. When more inequalities are used, the physical region containing
the correct parameters reduces to the intersection of all physical
hyperspaces. Eventually, the region reduces to a small, convex cell (not
necessarily closed) whose walls are determined by a number of inequalities of
the order of the dimension of parameter space.

The optimal point in the cell is found by using 
perceptron strategy, as described in Section \ref{methods}. This
procedure has been shown to converge to the true potential when used
in exact models where rigorous test are available \cite{VC98}
\cite{CM98}. It is also possible that parametrizations (\ref{nrgns})
or (\ref{nrgs}) may not be sufficient to guarantee that a solution to
inequalities~(\ref{ineq}) exists. Indeed, if the decoys structures are
very competitive with the native structures, three or further body
interactions might be necessary to solve inequality (\ref{ineq})
consistently \cite{VN99}.

This procedure differs significantly from the one of
\citeasnoun{MC92} where the parameters were determined in a sub-optimal manner.

\section{Results and discussion}
\label{sec:res}

We succeeded in finding an optimal solution to the different systems
of inequalities (\ref{ineq}): the optimal parameters obtained for
Hamiltonians (\ref{nrgs}) and $f=1$ are given in table
\ref{parameters}.

We found that only a tiny fraction of all inequalities (\ref{ineq}) determine
the optimal stability solution (more or less 100 out of 1551196 according to
the $f$ used or whether solvation term is present).  It is  important to ensure
that the optimal solution does not fluctuate wildly when stringent inequalities
are added or removed. To check this, we eliminated the 100 most  stringent
inequalities. Even though this completely replaces the walls of the physical
cell, the new optimal solution slightly differed from the first one:
representing the parameters in a 230-dimensional vector space the two vectors
were only $15^o$ apart \footnote{We note that only the direction of the vector
of parameters is important, because it sets the rank of the conformations that
a sequence can assume, while the norm of that vector just sets an energy scale}.
Such a degree of correlation is significant because the expected angle
between two uncorrelated vectors in a space of $\approx 200$ dimensions is
about $90^o \pm 4^o$. This gives confidence in the robustness of the procedure
and the statistics of the training set.

The optimal parameters extracted with different trial forms of $f$ in
(\ref{ineq}) were also closely correlated. As summarized in table
\ref{tvangle}, their relative angle was always less than $15^o$. On
the contrary, sub-optimal vectors, for which inequalities (\ref{ineq})
are satisfied for $c \approx 0$ (in which case the detailed form of
$f$ is not relevant) form, on average, an angle of $50^0$ with the
optimal solution. This fact underscores the importance of introducing an
extremal criterion when maximizing (\ref{ineq}).

\begin{figure}
\begin{center}
\epsfig{figure=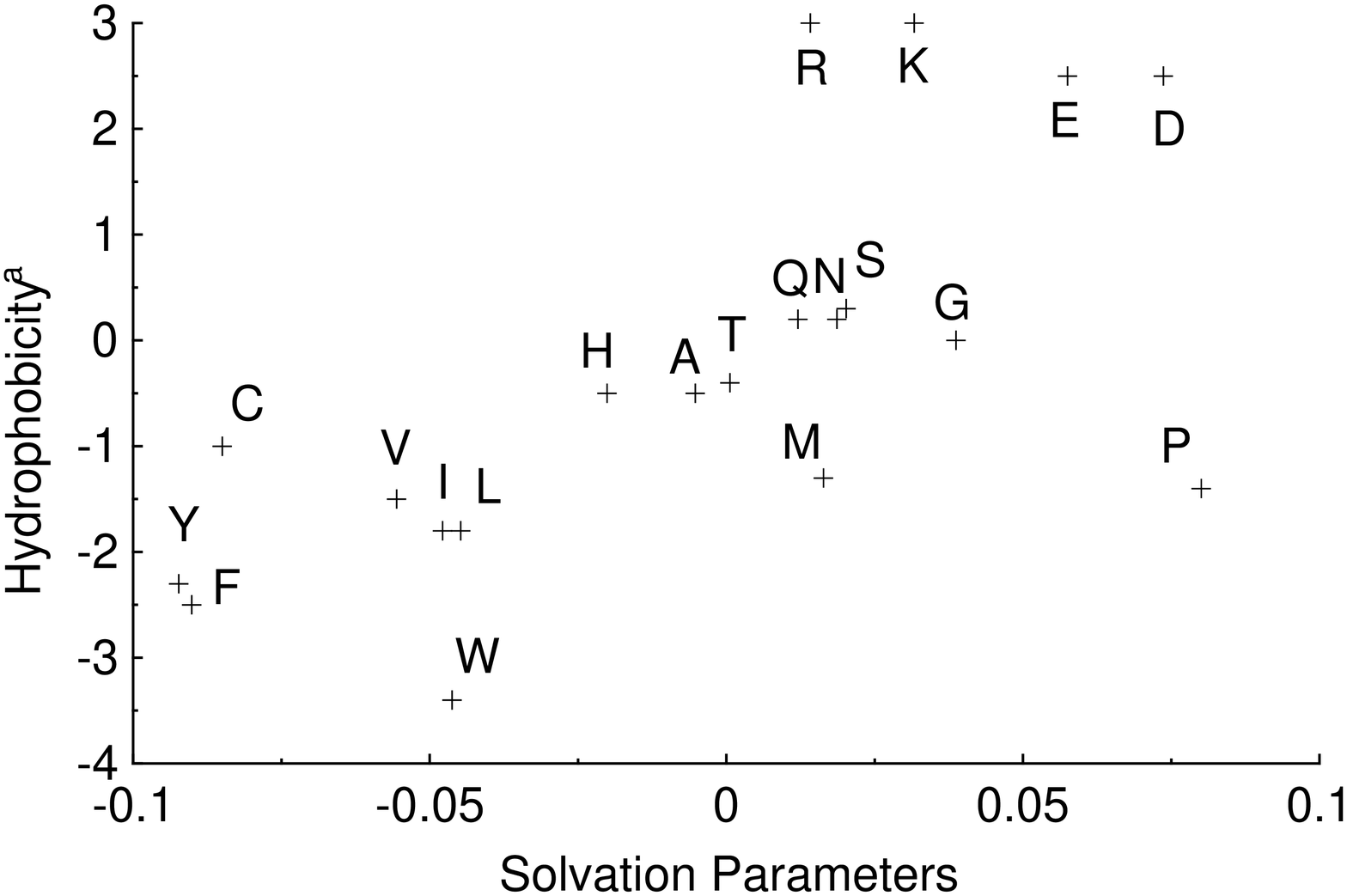,height=5cm,angle=0}
\label{hydroph}
\caption{ The extracted solvation parameters, $\epsilon(0,A)$
(see eqn.~(\protect{\ref{nrgs}})) versus standard hydrophobicity values
(Creighton 1993, p.154). }
\end{center}
\end{figure}

The extracted solvation parameters showed a very good correlation (0.67
correlation coefficient) with
the hydrophobicity scales as given by \citeasnoun{CRE}. As shown in
Fig. \ref{hydroph}, the agreement is quite good except, perhaps, for
proline. The discrepancy with proline finds a natural explanation within
the scheme that we used. In fact, while the hydrophobicity scales in
Fig.~\ref{hydroph} relate to the propensities of individual, isolated
amino acids, the solvation parameter reflects also their structural
functionality in a peptide context. In fact, because the prolines are
typically located in loop regions, they appear to have an effective
hydrophilic propensity larger than their bare value.

Finally, we carried out a stringent validation of the extracted potentials by
performing a blind ground-state recognition on a test set. The test set (see
Table \ref{maintable})was comprised of proteins taken from those used in 
\citeasnoun{MJ96} and chosen so that they would meet some of the criteria used
to select the training set (see section \ref{methods}). We deliberately
introduced proteins with hetero groups, low degree of compactness and also
pairs with high structural homology. In all cases we ensured that no protein in
the test set had a significant degree of structural homology with those in the
training one.

We took, in turn, the sequences of the test set and threaded them
on structures in the set with equal or longer length. Hence, we checked
whether using the optimal potential parameters of Table \ref{parameters}, the
true native state was recognized as the lowest energy one. Indeed,
this turned out to be the case for all but 6 proteins. No higher
success rate was found on using some other known sets of potentials
consistent with the form of our Hamiltonian.

Another relevant quantity related to the performance of the algorithm is given
by the number of wrongly satisfied inequalities of type (\ref{ineq}) for the
test set. This quantity shows a much higher degree of variability than the
number of correctly identified ground states and is given in column 3 of Table
\ref{successr}. It can be seen that the optimal parameters extracted with the
solvent and $f=1$ perform far better than those without the solvent and
previously extracted ones. It also appears that, enforcing optimality provides
a dramatic reduction of wrong inequalities compared to the sub-optimal cases.
This provides a sound {\em a posteriori} justification for the optimal
extraction procedure as well as giving confidence in the parameters.

\begin{figure}
\begin{center}
\epsfig{figure=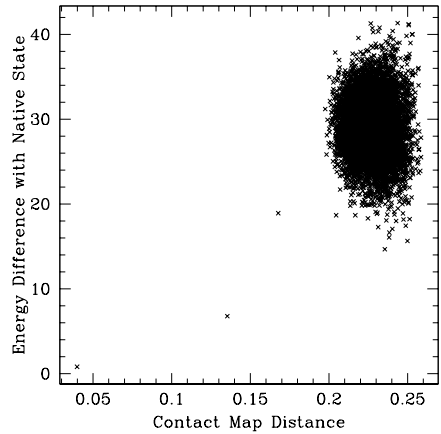,height=6cm}
\end{center}
\caption{\label{energylandscape} Energies of the protein sequence 1lz3
when threaded on decoy structures against structual dissimilarity. Very low
energies are observed when threading on homologous conformations (in
particular protein 132l).}
\end{figure}

The few cases where the extracted potentials fail are due to one of
the following situations: a) the native protein is not too compact or
b) it contains stabilizing hetero groups. Situations in which a highly
homologous structure has a lower energy score than the native one are
not deemed as errors. A typical energy/structural distance plot is
shown in Fig.~\ref{energylandscape}.  It is apparent that homologous
structures have energies similar to that of native conformations, while
distant structures lie higher in energy. Some differences in the performance
were observed for the sets of 210 and 230 potentials. While the
latter only fail to recognize native states containing heme groups
etc., the former occasionally fail to recognize the native states with no
atypical feature (e.g. interleukin-4, 1rcb).

For proteins with heme groups, several structures score
better: they usually present a smaller number of contacts than
the native structure being less compact than the native state.
This is possibly related to the presence of
proline in an unusually buried position, namely the heme pocket. In
fact, due to the high effective solvation term assigned to proline,
the native structure is penalized with respect to decoy ones where it
is confined in more solvent-exposed positions.

An interesting case where the failure relates to a non-compact protein, is
given by {\em trp}~aporepressor (3wrp), for which several better scoring decoys exist.
The explanation lies in the fact that 3wrp is always found as a dimer: the side
of the protein binding its counterpart has non-polar surface residues usually
in contact with non-polar residues on the other dimer, which is not accounted
for by our procedure.

Nevertheless, the algorithm appears to work in other instances of non-compact
conformations such as troponin~{\em c} (4tnc) and calmodulin (1cll) and on some
cytochrome-{\em c} as 1ccr or 1yeb, showing that the optimization procedure
succeeds in extracting a potential with a wider applicability range than that given
by the folds used in the training set.

Over 15 different pairs of homologous structures (contact map distance
less than 0.1), the energy functional is able to rank the true native
state as the lowest in just 8 cases. In the other cases the native
state attains an energy value slightly higher than the homologous one. As
expected, the simple contact potential cannot distinguish the native
state among very similar structures but it consistently assigns
similar value of energy to similar conformations according to the
degree of similarity (see Fig.~\ref{energylandscape}).

\begin{figure}
\begin{center}
\epsfig{figure=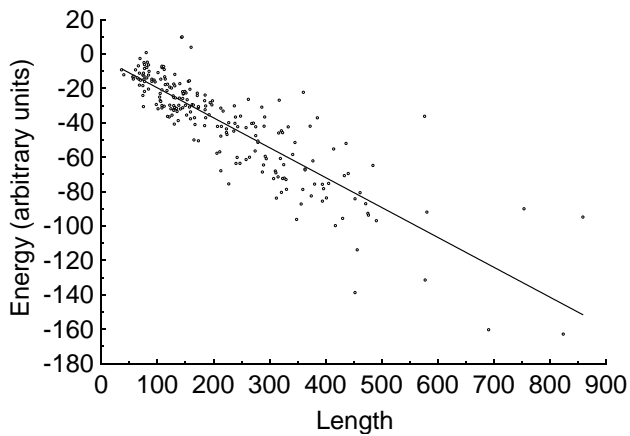,height=6cm,angle=0}
\end{center}
\caption{\label{energyvsLength} When using the extracted parameters of
Table \ref{parameters} the native state energy of proteins shows an approximately
linear behavious as a function of their length. Points refer to
proteins for both the training and test sets. Proteins with less than
200 amino acids and atypical compactness present significant deviations from the
average trend.}
\end{figure}

It is important to note that there is a well-defined trend for the
protein ground-state energies as a function of protein
length. Deviations from this typical trend could be used to assess the
reliability of the predicted fold of a sequence with unknown structure
(Fig.~\ref{energyvsLength}).

\section{Methods} 
\label{methods}

\subsection{Protein data sets} 

We selected 142 protein structures from PDB \cite{PDB}, listed in
Tab.~\ref{training.table}, with lengths varying from 36 to 823,
following criteria very similar to \citeasnoun{MC92}. For each
reference protein, we built a set of alternative conformations by
threading its sequence on all the other structures in the set with a
greater or equal number of amino acids. As explained in
\citeasnoun{JT92}, threading a sequence of $L$ amino acids on a
structure, $\Gamma^\prime$, of length $L^\prime>L$, involves mounting
the sequence on all the ($L^\prime-L+1$) segments (of contiguous amino
acids) taken from $\Gamma^\prime$. This procedure assigns the contact
map of the threaded segment to the threading sequence. The inherited
contact map is used to calculate the energy of the sequence in the
alternative, threaded, conformation, and compared with the energy in
its native state, which is required  to be the global energy minimum.

Only single chain structures have been selected in order to avoid the
occurrence of interchain contacts between amino acids, that are not
detected by our procedure and that could cause the stabilization of
hydrophobic residues on a protein's surface. Considering multiple
chain structures would have introduced spurious effects in the
extracted potentials, since inter-chain contacts would not be present
in threaded conformations. For simplicity, however, we decided to
retain proteins which may be found in polymeric forms.
Because the presence of large hetero groups can distort the usual
geometry of dihedral angles between amino acids and cannot be treated
in a simple way by a pairwise potential, we discarded protein
structures with high percentages of non-water HETATM records in their
PDB files (like HEM or CPS groups).

We used the classification of 3-D protein structures SCOP \cite{MB95} to select
proteins spanning a wide range of different three dimensional folds: no pair of
proteins in our training set belong to the same SCOP family. Furthermore, we
have included only proteins in the first 4 SCOP classes: all-$\alpha$,
all-$\beta$, $\alpha$/$\beta$ and $\alpha + \beta$ proteins. Cell membrane or
surface proteins and very short peptide chains are excluded because usually
they are not stabilized by just amino acid interactions but by some external
factors, such as the hydrophobic environment, metal ligands, heme groups etc. No
unresolved backbone atoms inside a chain are allowed; disordered or unresolved
terminal backbone atoms are eliminated.

We also disregarded proteins that were not typically compact: because there is
a clear dependence of the radius of gyration and the number of contacts among
amino acids on chain length, we rejected from our training set proteins with
too large a radius of gyration or with significantly fewer contacts than expected for
their length. The rejection was based using the quantitative procedures
discussed in \citeasnoun{MC92}.

\subsection{Optimal Stability Perceptron}

It is convenient to recast expression (\ref{ineq}) so that the
dependence from the interaction parameters between amino acid types
$A$ and $A^\prime$, $\epsilon(A, A^\prime)$ appears explicitly,

\eqa{\label{exp11}\sum_{\it{A} \leq A^{\prime}} \frac{\epsilon(A,A^{\prime}
)}{|\vec{\epsilon}|} \cdot
\frac{n_{S_i,\Gamma_D}(A,A^{\prime} )- n_{S,\Gamma_i}(A,A^{\prime} )}{
f\left(D(\Gamma_D,\Gamma_i)\right)} &>& c \ .}

\noindent where $n_{S,\Gamma}(A,A^{\prime} )$ is the number of
contacts between amino acids $A$ and $A^\prime$ attained by $S$ on
$\Gamma$. The indices $A$ and $A^\prime$ run over the 20 amino acid
classes for parametrization \ref{nrgns}. 
Expression \ref{exp11} can be rewritten in a more compact form by
mapping the independent entries of the $\epsilon$ matrix on a
one-dimensional vector,

\begin{equation}
\vec{\epsilon} \equiv \{ \epsilon(1,1), \epsilon(1,2), ... ,
\epsilon(20,20)\} \ .
\end{equation}

\noindent and likewise for the vector
\eq{{\vec{N}_{S,\Gamma,\Gamma^\prime} \equiv
\{\frac{(n_{S,\Gamma}(1,1)-n_{S,\Gamma^\prime}(1,1))}{f\left(D(\Gamma^\prime,\Gamma)\right)} , ...  \}/.}}  With
the former definitions, equation (11) becomes:

\begin{equation}
\frac{\vec{\epsilon}}{|\vec{\epsilon}|} \cdot \vec{N}_{S_i, \Gamma_i, \Gamma_D} > c \ .
\label{eqn:simple}
\end{equation}

A formally equivalent expression is obtained when using 230 parameters
as in eq. (\ref{nrgs}).

\noindent Expression (\ref{eqn:simple}) leads to a geometrically
appealing interpretation of the stability requirement. The optimal
stability is reached when the interaction vector has the largest
possible inner product with all the $\vec{N}_{S_i,\Gamma_i,\Gamma_D}$
vectors, also termed ``patterns'', originated from the training set. A
rigorous solution for this geometrical problem was given by
\citeasnoun{KM87} who suggested an iterative procedure called optimal
stability perceptron.

\begin{figure}
\begin{center}
\epsfig{figure=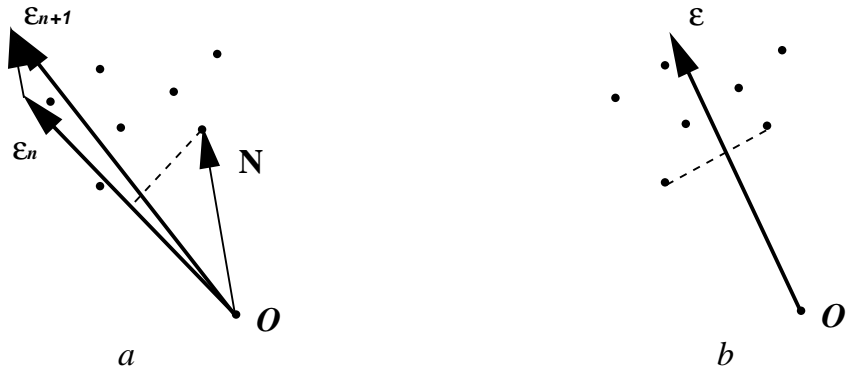,height=5cm}
\end{center}
\caption{\label{perceptron} Schematic representation of a typical
perceptron update in a two-dimensional space. Inequalities are
represented by vectors connecting the origin to the points. At
ineration $n$, The stability, $c$, of ineq. (\protect{\ref{ineq}}) is the
smallest inner product between the parameter vector $\vec{\epsilon}_n$
and each of the inequalities. In case {\em (a)} $c$ is given by
$\vec{\epsilon} \cdot \vec{N}$ and $\vec{\epsilon}_n$ accordingly
aquires a small component parallel to $\vec{N}$. An equilibrium
situations is shown in {\em (b)}. Successive updates cause
$\vec{\epsilon}$ to bounce on either side of the equilibrium
direction. The latter is reached in a finite time, because the
relative size of the added component decreases with the number of
iterations.}
\end{figure}

The procedure is the following. Starting from a random (or an
otherwise assigned) set of interactions satisfying the norm constraint
(\ref{eqn:norm}), the stability score of all inequalities is
computed. Then, the potentials are updated so to increase the
stability of the lowest scoring inequality. This is done by adding to
the original potentials vector, $\vec{\epsilon}$, a small term
proportional to the worst scoring pattern (see Fig. \ref{perceptron}):

\eq{\forall A ,A^{\prime} \quad \epsilon (A ,A^{\prime} ) \rightarrow 
\epsilon(A,A^{\prime})+
\frac{1}{d} N_{S,\Gamma,\Gamma^\prime}(A,A^{\prime})\ ,
\label{eqn:upd}}

\noindent where $d$ is the dimension of the parameter space (210 or
230). Then, the inequalities are re-computed with the updated
interaction parameters. The lowest scoring one is identified again and
a new update of $\vec{\epsilon}$ is carried out. Note that the update
(\ref{eqn:upd}) will typically change the norm of $\vec{\epsilon}$. 
The unit norm (see eqn. (\ref{eqn:norm})) can be conveniently enforced
after convergence has been achieved.

While convergence is guaranteed to be reached in a finite number of
steps, the time required for each iteration grows linearly with the
number of inequalities. In our case, we typically dealt with $\approx
10^6$ inequalities, and convergence sometimes  required several thousand
iterations (each taking few seconds of CPU time). Hence, we devised a
scheme to speed up convergence based on the observation that the
stability variation due to the change of parameters is proportional to
the distance between the inequality point and the parameter direction,

\eqa{\Delta s= & \vec{N} \cdot \Delta \vec{\epsilon}= \vec{N}_{\perp
\vec{\epsilon}} \cdot \Delta \vec{\epsilon}_{\perp
\vec{\epsilon}}+\underbrace{\vec{N}_{\parallel \vec{\epsilon}} \cdot
\Delta \vec{\epsilon}_{\parallel \vec{\epsilon}}}_{\ge 0.0}\propto
\nonumber \\ & \propto  \underbrace{a}_{\not= 0.0}\cdot |\vec{N}_{\perp
\vec{\epsilon}}|+\underbrace{b}_{\ge 0.0} &\ .}

\begin{figure}
\begin{center}
\epsfig{figure=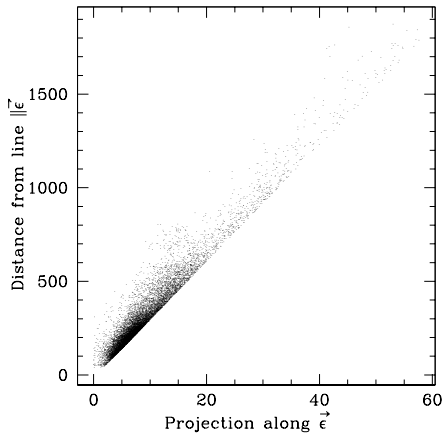,height=6cm}
\end{center}
\caption{\label{dropineq} After a sufficient number of iterations,
successive perceptron updates will not change 
appreciably $\vec{\epsilon}$. To speed
up convergence, it is convenient to temporarily retain only those
inequalities (points in parameter space) lying outside a cone with
axis along $\vec{\epsilon}$ and suitable vertex and width (see
text). 
The edge of the cone is visible in this figure, where only the
retained inequalities are shown.}
\end{figure}

This implies that inequality vectors far from the parameter direction
will get the largest score variations (positive or negative) and so
they are more likely to become the lowest scoring ones. Accordingly,
the standard perceptron procedure was run until reaching a sub-optimal
value for the stability threshold, $c>0$; this typically needed 300
iterations. Then we temporarily restricted the updating procedure to
those inequalities lying outside a cone with axis along the parameter
direction and vertex at distance larger than $c$ from origin (see
Fig. \ref{dropineq} ). The cone width was determined to limit the
number of inequalities to less than $20000$ (Fig. \ref{dropineq}). In
this way we had to deal with $10^4$ inequalities that are 2 orders of
magnitude smaller than the original ones, thus decreasing enormously
the CPU time needed for optimization. Furthermore, after convergence,
the neglected inequalities are found to be satisfied well above the
optimal stability threshold $c_{max}$, thus justifying the numerical
shortcut.

\begin{figure}
\begin{center}
\epsfig{figure=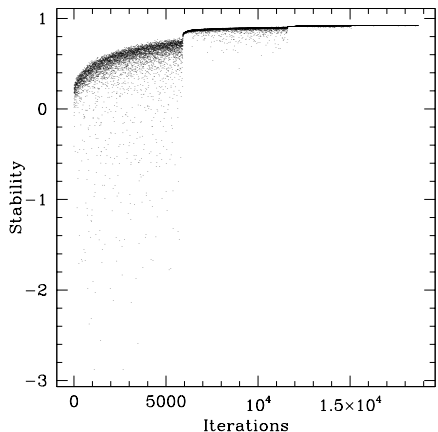,height=6cm}
\caption{\label{badconv} Perceptron stability as a function of the
number of iterations. The discontinuities are associated with the
``inflation'' of $\vec{\epsilon}$ used to speed up convergence (see
text).}
\end{center}
\end{figure}

We conclude by remarking that, if the relative correction to
$\epsilon(A,A^\prime)$ in eq. (\ref{eqn:upd}) is too large, this may
result in a slowing down of the convergence. This difficulty can be
readily circumvented by increasing the size of $\vec{\epsilon}$ by an
order of magnitude. It was typically necessary to repeat this
``inflation'' procedure 3-4 times during each run towards
convergence (see Fig. \ref{badconv}). This was sufficient to reach
optimal convergence to the solution: $\Delta c/c < 10^-3$.
\vskip 1.0cm
\noindent{\bf Acknowledgments.} We acknowledge support from 
INFM, NASA, NATO and the Petroleum
Research Fund administered by the American Chemical Society. We thank  
C. Clementi, R. Dima, L. Guidoni, S. Piana, A. Rossi, F. Seno and J.~Van~Mourik for useful
discussions.

\onecolumn
\begin{table}
\begin{center}
\rotatebox[origin=bl]{270}{
\scalebox{0.5}{\begin{tabular}{||c|c|c|c|c|c|c|c|c|c|c|c|c|c|c|c|c|c|c|c|c||}
\cline{1-21}
A & -0.0269 & & & & & & & & & & & & & & & & & & & \\ 
C & -0.0142 & -0.1509 & & & & & & & & & & & & & & & & & & \\ 
D & 0.0222 & 0.0027 & -0.0130 & & & & & & & & & & & & & & & & & \\ 
E & 0.0308 & 0.0047 & 0.1587 & 0.0712 & & & & & & & & & & & & & & & & \\ 
F & 0.1415 & -0.0500 & 0.0167 & 0.0458 & -0.1098 & & & & & & & & & & & & & & &
\\ 
G & 0.0386 & 0.0635 & -0.0015 & 0.0205 & -0.0774 & 0.0023 & & & & & & & & & & &
& & & \\ 
H & 0.0392 & -0.0293 & 0.0255 & -0.0472 & 0.0240 & -0.0322 & 0.0046 & & & & & &
& & & & & & & \\ 
I & -0.0261 & -0.0960 & 0.1188 & 0.1026 & -0.0423 & 0.0310 & 0.0175 & -0.1753 &
& & & & & & & & & & & \\ 
K & 0.0760 & 0.0372 & -0.0675 & -0.2113 & 0.0302 & -0.0050 & 0.0137 & 0.1177 &
0.0498 & & & & & & & & & & & \\ 
L & -0.0540 & 0.0776 & 0.0078 & 0.0740 & -0.0988 & -0.0787 & 0.0849 & -0.0672 &
-0.0431 & -0.1246 & & & & & & & & & & \\ 
M & 0.0070 & 0.0239 & -0.0995 & 0.0116 & -0.0614 & -0.0241 & -0.0603 & -0.0472 &
0.0807 & 0.0179 & 0.0393 & & & & & & & & & \\ 
N & 0.0036 & -0.0158 & 0.0208 & -0.1343 & 0.1070 & 0.0377 & -0.0005 & 0.1172 &
-0.0636 & 0.1172 & -0.0070 & -0.0064 & & & & & & & & \\ 
P & -0.0760 & 0.0134 & 0.0180 & 0.0272 & 0.1145 & -0.0043 & 0.0615 & 0.0253 &
-0.0798 & 0.0125 & -0.0205 & 0.0713 & 0.0381 & & & & & & & \\ 
Q & -0.1203 & 0.0294 & 0.0724 & 0.0274 & 0.0853 & 0.0527 & -0.0068 & 0.0530 &
-0.0691 & -0.0792 & 0.0176 & -0.0199 & -0.0460 & 0.0141 & & & & & & \\ 
R & 0.0154 & 0.0202 & -0.1486 & -0.1610 & -0.1207 & -0.0010 & -0.0315 & 0.0528 &
0.0892 & -0.1136 & 0.0646 & 0.0225 & -0.0228 & 0.0233 & 0.0624 & & & & & \\ 
S & 0.0358 & 0.0047 & -0.1300 & -0.0895 & 0.0580 & -0.0257 & -0.0469 & 0.0288 &
0.0214 & 0.2016 & -0.0236 & -0.1070 & 0.0085 & 0.0671 & 0.0733 & -0.0618 & & & &
\\ 
T & -0.0831 & -0.0511 & -0.0687 & 0.1258 & -0.0492 & 0.0526 & -0.0890 & -0.0541
& 0.0180 & 0.1139 & 0.0914 & -0.1003 & 0.0623 & 0.0087 & 0.0340 & -0.0576 &
0.0138 & & & \\ 
V & -0.0658 & -0.0066 & 0.0702 & -0.0609 & -0.0185 & -0.0406 & 0.0911 & -0.0908
& -0.0146 & -0.0724 & 0.0465 & 0.0365 & -0.0351 & -0.0155 & 0.0528 & 0.0927 &
0.0414 & -0.0637 & & \\ 
W & 0.0878 & 0.0129 & 0.0575 & 0.0093 & 0.0210 & 0.0055 & 0.0123 & -0.0631 &
0.0281 & 0.0058 & 0.0380 & -0.0487 & -0.0345 & -0.0733 & 0.0007 & -0.0076 &
0.0234 & -0.0504 & 0.0118 & \\ 
Y & -0.0367 & 0.0389 & 0.0111 & 0.0521 & -0.1059 & 0.0249 & -0.0507 & -0.0505 &
0.0235 & -0.0265 & -0.0785 & -0.0117 & -0.0536 & -0.0089 & 0.1021 & -0.0219 &
-0.0317 & 0.0482 & -0.0826 & 0.1660  \\ 
Sol & -0.0053 & -0.0850 & 0.0737 & 0.0575 & -0.0901 & 0.0387 & -0.0201 & -0.0478
& 0.0317 & -0.0448 & 0.0164 & 0.0186 & 0.0801 & 0.0121 & 0.0141 & 0.0202 &
0.0006 & -0.0556 & -0.0462 & -0.0923 \\ 
\cline{1-21}
& A & C & D & E & F & G & H & I & K & L & M & N & P & Q & R & S & T & V & W & Y
\\ 
\cline{1-21}
\end{tabular}
}
}
\end{center}
\caption{\label{parameters} Interaction parameters extracted
by using energy funtional ~(\ref{nrgs}).}
\end{table}
\newpage

\begin{table}
\begin{center}
\begin{tabular}{||c|c||c||}
\hline
\multicolumn{3}{|c|}{Solvent}\\
\hline
$f=1$ &$f=x$ & $10.1^o $\\
$f=1$ &$f=x^2$ & $13.5^o$\\
$f=x$ & $f=x^2$ &  $6.67^o $\\
$f=1$ &Non-optimal  &(average) $54^o$\\
$f=x$ &Non-optimal &(average) $55^o $\\
$f=x^2$ &Non-optimal & (average) $55^o $\\
Non-optimal &Non-optimal &(average) $67^o$\\
\hline
\multicolumn{3}{c|}{ No-Solvent} \\
\hline
$f=1$ &$f=x$ & $8.5^o$\\
$f=1$ &$f=x^2$ & $15.9^o$\\
$f=x$ & $f=x^2$ &  $10.0^o$\\
\hline
\end{tabular}
\end{center}
\caption{Angles formed by the optimal vectors for various forms of the
Hamiltonian and $f$ (see eqns. (\ref{nrgns}), (\ref{nrgs}) and
(\ref{ineq}). \label{tvangle}}
\end{table}
\newpage

\begin{table}
\begin{center}
\scalebox{0.5}[0.5]{\begin{tabular}{||c||c|c|c|c|c|c|c||}
\hline
Prot. code & Length & Native Energy & No. Decoys & No. Better Str. & Average  & \multicolumn{2}{|c||}{SCOP}\\
           &        &               &            &                 & Difference & \multicolumn{2}{|c||}{Classification}\\
           &				&								&							&									& in Contacts &\multicolumn{2}{|c||}{ }\\
\hline
1hbg & 146 & -23.259936 & 8007 & 0 & 0  $\pm$  0 & 1001001001001 & 003 \\
1mba & 146 & -22.890319 & 8007 & 0 & 0  $\pm$  0 & 1001001001001 & 005 \\
1mbs & 151 & -22.674067 & 7756 & 0 & 0  $\pm$  0 & 1001001001001 & 006 \\
1lh1 & 152 & -35.490600 & 7707 & 0 & 0  $\pm$  0 & 1001001001001 & 014 \\
2lhb & 149 & -21.766920 & 7855 & 0 & 0  $\pm$  0 & 1001001001001 & 033 \\
1cty & 108 & -7.089541 & 10259 & 5 & -57  $\pm$  81 & 1001003001001 & 004 \\
1yeb & 108 & -7.110300 & 10259 & $1^\dagger$ & -3  $\pm$  0 & 1001003001001 & 004 \\
1ccr & 111 & -7.226756 & 10056 & 0 & 0  $\pm$  0 & 1001003001001 & 006 \\
2c2c & 112 & -8.985422 & 9989 & 2 & -150  $\pm$  170 & 1001003001001 & 009 \\
351c & 82 & -4.752413 & 12127 & 3 & 17  $\pm$  50 & 1001003001001 & 017 \\
$\star$ 1le4 & 139 & -25.618122 & 8383 & 0 & 0  $\pm$  0 & 1001023001001 & 003 \\
2mhr & 117 & -21.248559 & 9670 & 0 & 0  $\pm$  0 & 1001023004001 & 004 \\
1rcb & 129 & -30.026804 & 8944 & 0 & 0  $\pm$  0 & 1001025001002 & 002 \\
$\star$ 4tnc & 160 & 3.959946 & 7332 & 0 & 0  $\pm$  0 & 1001034001005 & 001 \\
$\star$ 1cll & 143 & 9.773591 & 8166 & 0 & 0  $\pm$  0 & 1001034001005 & 005 \\
$\star$ 1clm & 144 & 9.950750 & 8112 & 1 & -370  $\pm$  0 & 1001034001005 & 011 \\
1cca & 291 & -36.626670 & 2654 & 0 & 0  $\pm$  0 & 1001065001001 & 003 \\
$\star$ 3wrp & 101 & -10.772739 & 10755 & 9 & 235  $\pm$  90 & 1001078001001 & 001 \\
1poc & 134 & -31.651214 & 8659 & 0 & 0  $\pm$  0 & 1001095001001 & 001 \\
\hline
2imm & 114 & -17.753696 & 9858 & 0 & 0  $\pm$  0 & 1002001001001 & 024 \\
2rhe & 114 & -14.557044 & 9858 & 0 & 0  $\pm$  0 & 1002001001001 & 088 \\
2stv & 184 & -27.148904 & 6318 & 0 & 0  $\pm$  0 & 1002008001002 & 002 \\
2cna & 237 & -30.109204 & 4302 & 0 & 0  $\pm$  0 & 1002019001001 & 001 \\
1lec & 242 & -60.150421 & 4129 & 0 & 0  $\pm$  0 & 1002019001001 & 004 \\
1lte & 239 & -36.671976 & 4231 & 0 & 0  $\pm$  0 & 1002019001001 & 005 \\
1shg & 57 & -14.578324 & 14051 & 0 & 0  $\pm$  0 & 1002021002001 & 006 \\
8adh & 374 & -82.544846 & 1113 & 0 & 0  $\pm$  0 & 1002022001002 & 001 \\
1gbt & 223 & -43.532366 & 4821 & 0 & 0  $\pm$  0 & 1002031001002 & 001 \\
1est & 240 & -63.644775 & 4196 & 0 & 0  $\pm$  0 & 1002031001002 & 013 \\
4ape & 330 & -78.686637 & 1763 & 0 & 0  $\pm$  0 & 1002034001002 & 001 \\
3app & 323 & -72.409291 & 1897 & 0 & 0  $\pm$  0 & 1002034001002 & 002 \\
2apr & 325 & -64.356790 & 1856 & 0 & 0  $\pm$  0 & 1002034001002 & 003 \\
3pep & 326 & -72.156108 & 1836 & 0 & 0  $\pm$  0 & 1002034001002 & 006 \\
1mpp & 356 & -87.210810 & 1356 & 0 & 0  $\pm$  0 & 1002034001002 & 009 \\
1cms & 321 & -80.466445 & 1940 & 0 & 0  $\pm$  0 & 1002034001002 & 011 \\
1brp & 173 & -30.565982 & 6770 & 0 & 0  $\pm$  0 & 1002041001001 & 002 \\
1mup & 157 & -30.854982 & 7471 & 0 & 0  $\pm$  0 & 1002041001001 & 008 \\
2aaa & 475 & -92.682878 & 340 & $1^\dagger$ & 37  $\pm$  0 & 1002048001001 & 008 \\
6taa & 476 & -93.727839 & 335 & 0 & 0  $\pm$  0 & 1002048001001 & 009 \\
\hline
1btc & 490 & -96.939667 & 292 & 0 & 0  $\pm$  0 & 1003001001002 & 001 \\
1ald & 363 & -67.400922 & 1257 & 0 & 0  $\pm$  0 & 1003001003001 & 002 \\
3enl & 436 & -52.025765 & 553 & 0 & 0  $\pm$  0 & 1003001006001 & 001 \\
1pii & 452 & -138.741679 & 456 & 0 & 0  $\pm$  0 & 1003001008001 & 001 \\
1xis & 385 & -37.102765 & 980 & 0 & 0  $\pm$  0 & 1003001012001 & 004 \\
1phh & 394 & -85.545211 & 880 & 0 & 0  $\pm$  0 & 1003004001002 & 002 \\
1gal & 580 & -91.922051 & 111 & 0 & 0  $\pm$  0 & 1003004001002 & 004 \\
1dhr & 236 & -44.849628 & 4339 & 0 & 0  $\pm$  0 & 1003019001002 & 006 \\
2cmd & 312 & -83.791376 & 2147 & 0 & 0  $\pm$  0 & 1003019001005 & 002 \\
1ldm & 329 & -73.012055 & 1781 & 0 & 0  $\pm$  0 & 1003019001005 & 008 \\
1gky & 186 & -30.194166 & 6237 & 0 & 0  $\pm$  0 & 1003025001001 & 001 \\
3adk & 194 & -28.274731 & 5924 & 0 & 0  $\pm$  0 & 1003025001001 & 006 \\
121p & 166 & -37.077357 & 7067 & $1^\dagger$ & 84  $\pm$  0 & 1003025001003 & 001 \\
4q21 & 167 & -40.978706 & 7023 & 0 & 0  $\pm$  0 & 1003025001003 & 001 \\
1sbc & 274 & -56.392841 & 3113 & 0 & 0  $\pm$  0 & 1003028001001 & 001 \\
1thm & 279 & -51.381279 & 2968 & 0 & 0  $\pm$  0 & 1003028001001 & 003 \\
1s01 & 275 & -47.482543 & 3082 & 0 & 0  $\pm$  0 & 1003028001001 & 006 \\
1s02 & 275 & -42.291256 & 3082 & $1^\dagger$ & 3  $\pm$  0 & 1003028001001 & 006 \\
2prk & 279 & -50.558910 & 2968 & 0 & 0  $\pm$  0 & 1003028001001 & 007 \\
1ama & 401 & -83.687554 & 813 & 0 & 0  $\pm$  0 & 1003048001001 & 001 \\
1spa & 396 & -78.338515 & 859 & 0 & 0  $\pm$  0 & 1003048001001 & 004 \\
1ipd & 345 & -51.476690 & 1522 & 0 & 0  $\pm$  0 & 1003057001001 & 001 \\
3icd & 414 & -55.210210 & 708 & 0 & 0  $\pm$  0 & 1003057001001 & 003 \\
1rhd & 292 & -59.479630 & 2628 & 0 & 0  $\pm$  0 & 1003060001001 & 001 \\
3pfk & 319 & -60.742159 & 1985 & 0 & 0  $\pm$  0 & 1003070001001 & 002 \\
1ovb & 159 & -37.670779 & 7378 & 0 & 0  $\pm$  0 & 1003073001002 & 002 \\
1lfg & 690 & -160.321444 & 0 & 0 & 0  $\pm$  0 & 1003073001002 & 005 \\
\hline
132l & 129 & -32.386126 & 8944 & $1^\dagger$ & 14  $\pm$  0 & 1004002001002 & 001 \\
1lz3 & 129 & -32.719245 & 8944 & 0 & 0  $\pm$  0 & 1004002001002 & 002 \\
1laa & 130 & -40.734890 & 8884 & 0 & 0  $\pm$  0 & 1004002001002 & 008 \\
1alc & 121 & -36.611850 & 9425 & 0 & 0  $\pm$  0 & 1004002001002 & 013 \\
3il8 & 68 & -18.129049 & 13192 & 0 & 0  $\pm$  0 & 1004007001001 & 001 \\
1fkb & 106 & -14.625747 & 10399 & 0 & 0  $\pm$  0 & 1004019001001 & 001 \\
1yat & 113 & -13.260539 & 9923 & 0 & 0  $\pm$  0 & 1004019001001 & 003 \\
1ctf & 68 & -11.121616 & 13192 & 0 & 0  $\pm$  0 & 1004026001001 & 001 \\
1fd2 & 106 & -27.581742 & 10399 & 0 & 0  $\pm$  0 & 1004033001002 & 001 \\
2fxb & 81 & -7.781913 & 12202 & 0 & 0  $\pm$  0 & 1004033001004 & 003 \\
3tms & 264 & -60.007310 & 3424 & 0 & 0  $\pm$  0 & 1004063001001 & 001 \\
3b5c & 84 & -12.241707 & 11980 & 0 & 0  $\pm$  0 & 1004066001001 & 001 \\
\hline
\end{tabular}
}
\end{center}
\caption{ \label{maintable} Proteins used in the test set. The symbol
$\dagger$ denotes instances where the better scoring structure is
homologous to the target protein, while a $\star$ marks non compact
native state. 1lfg has been used only as structural template.}
\end{table}

\begin{table}
\begin{tabular}{||c||c|c||}
\hline
& \multicolumn{2}{c||}{Score on Test}			\\
\hline
Solvent & Not Rec. Str.&		Unsat. Ineq.\\
\hline
$f=1$& 5 	& 25												\\
\hline
$f=x$	& 5 & 33													\\
\hline
$f=x^2$	& 5 & 36													\\
\hline
Non-Optimal& 5.25 & 75.7																	\\
\hline
No Solvent & & 																				\\
\hline
$f=1$& 6 & 118													\\
\hline
$f=x$	& 6 & 200													\\
\hline
$f=x^2$	& 6 & 254												\\
\hline \hline
DBMS & 7  & 51\\
\hline
KGS & 6 & 452\\
\hline
MC &  6 & 48  \\
\hline
\hline
& \multicolumn{2}{c||}{Score on Training}			\\
\hline
DBMS &  13  & 1091 \\
KGS &  22 & 9789 \\
MC &  20& 1826 \\
\hline
\end{tabular}
\label{successr}
\caption{ Performance of the potentials extracted in this work
and other known sets. The second column gives the number of unrecognised native
states among the 78 ones of Table~\protect{\ref{training.table}}. The
associated number of violated inequalities (against a total of 444199) is given
in column 3. The acronyms for the alternative potentials refer to:
DBMS$=$(Dima et al. 1999),
KGS$=$(Kolinsky et al. 1993),
MC$=$(Maiorov \& Crippen 1994).
The last part of the table
shows the scores of the alternative potentials applied to our training set of
142 proteins generating 1551196 inequalities (on which  our potentials, by
definition, scores 100\% success). } 

\end{table}
\onecolumn 
\begin{longtable}{||c||c|c|c|c||}
\hline
Prot. Code& Length& \multicolumn{2}{|c|}{SCOP Classification}&No. Decoys\\
\hline
1vii & 36 & 1001014001001 & 001 & 25461 \\
1erd & 40 & 1001010001001 & 002 & 24896 \\
1pru & 56 & 1001030001003 & 001 & 22655 \\
1fxd & 58 & 1004033001001 & 001 & 22376 \\
1igd & 61 & 1004012001001 & 001 & 21961 \\
1orc & 64 & 1001030001002 & 005 & 21549 \\
1sap & 66 & 1004009001001 & 002 & 21276 \\
1mit & 67 & 1004022001001 & 003 & 21140 \\
1utg & 69 & 1001072001001 & 001 & 20871 \\
1ail & 70 & 1001015001001 & 001 & 20737 \\
1hoe & 74 & 1002004001001 & 001 & 20208 \\
1kjs & 74 & 1001040001001 & 001 & 20208 \\
1ubi & 74 & 1004012002001 & 001 & 20208 \\
1hyp & 75 & 1001042001001 & 001 & 20076 \\
5icb & 75 & 1001034001001 & 001 & 20076 \\
1fow & 76 & 1001004004001 & 001 & 19947 \\
1tif & 76 & 1004012006001 & 001 & 19947 \\
1tnt & 76 & 1001006001001 & 001 & 19947 \\
1acp & 77 & 1001026001001 & 001 & 19820 \\
1hdj & 77 & 1001002002001 & 001 & 19820 \\
1iba & 77 & 1004053001001 & 001 & 19820 \\
1vcc & 77 & 1004067001001 & 001 & 19820 \\
1coo & 81 & 1001032001001 & 001 & 19336 \\
1cei & 84 & 1001026002001 & 001 & 18978 \\
1ngr & 84 & 1001062001001 & 001 & 18978 \\
1opd & 85 & 1004052001001 & 003 & 18859 \\
1fna & 90 & 1002001002001 & 002 & 18278 \\
1hqi & 90 & 1004079001001 & 001 & 18278 \\
1who & 94 & 1002006003001 & 001 & 17820 \\
1pdr & 96 & 1002023001001 & 001 & 17593 \\
1beo & 98 & 1001096001001 & 001 & 17368 \\
1tul & 101 & 1002060004001 & 001 & 17034 \\
9rnt & 104 & 1004001001001 & 003 & 16703 \\
1aac & 105 & 1002005001001 & 001 & 16593 \\
1erv & 105 & 1003033001001 & 004 & 16593 \\
1jpc & 108 & 1002054001001 & 001 & 16270 \\
1kum & 108 & 1002003001001 & 005 & 16270 \\
1rro & 108 & 1001034001004 & 001 & 16270 \\
3ssi & 108 & 1004044001001 & 002 & 16270 \\
2mcm & 112 & 1002001006001 & 001 & 15854 \\
1mai & 118 & 1002037001001 & 001 & 15241 \\
1poa & 118 & 1001095001002 & 001 & 15241 \\
1whi & 122 & 1002025001001 & 001 & 14839 \\
1yua & 122 & 1004067001002 & 001 & 14839 \\
7rsa & 124 & 1004004001001 & 001 & 14641 \\
2phy & 125 & 1004061002001 & 001 & 14543 \\
1bfg & 126 & 1002028001001 & 001 & 14446 \\
3chy & 128 & 1003013002001 & 001 & 14255 \\
1pdo & 129 & 1003040001001 & 001 & 14160 \\
1tum & 129 & 1004062001001 & 001 & 14160 \\
1ifc & 131 & 1002041001002 & 002 & 13974 \\
1kuh & 131 & 1004050001001 & 001 & 13974 \\
1lis & 131 & 1001017001001 & 001 & 13974 \\
1rsy & 132 & 1002006001002 & 001 & 13882 \\
1cof & 135 & 1004060001002 & 001 & 13617 \\
2end & 137 & 1001016001001 & 001 & 13442 \\
5nul & 138 & 1003013004001 & 006 & 13355 \\
2sns & 140 & 1002026001001 & 001 & 13184 \\
1lcl & 141 & 1002019001003 & 004 & 13099 \\
1lba & 145 & 1004064001001 & 001 & 12766 \\
1pkp & 145 & 1004011001001 & 002 & 12766 \\
1vsd & 145 & 1003041003002 & 001 & 12766 \\
1npk & 150 & 1004033006001 & 002 & 12363 \\
1irp & 153 & 1002028001002 & 003 & 12125 \\
2rn2 & 155 & 1003041003001 & 001 & 11968 \\
1vhh & 157 & 1004034001002 & 001 & 11813 \\
1gpr & 158 & 1002059003001 & 001 & 11736 \\
1ra9 & 159 & 1003053001001 & 001 & 11660 \\
119l & 162 & 1004002001003 & 001 & 11437 \\
2cpl & 164 & 1002043001001 & 001 & 11290 \\
1sfe & 165 & 1001004002001 & 001 & 11217 \\
1wba & 171 & 1002028003001 & 001 & 10790 \\
2fha & 171 & 1001024001001 & 003 & 10790 \\
1amm & 173 & 1002009001001 & 001 & 10650 \\
2prd & 173 & 1002026005001 & 003 & 10650 \\
1ido & 184 & 1003045001001 & 002 & 9911 \\
153l & 185 & 1004002001004 & 001 & 9844 \\
1xnb & 185 & 1002019001008 & 001 & 9844 \\
1knb & 186 & 1002016001001 & 001 & 9778 \\
1kid & 192 & 1003005003001 & 001 & 9399 \\
1cex & 197 & 1003013007001 & 001 & 9088 \\
1chd & 198 & 1003027001001 & 001 & 9026 \\
1fua & 206 & 1003055001001 & 001 & 8545 \\
1thv & 207 & 1002018001001 & 001 & 8485 \\
2abk & 211 & 1001066001001 & 001 & 8252 \\
1ah6 & 213 & 1004068001001 & 001 & 8137 \\
1lbu & 213 & 1001019001001 & 001 & 8137 \\
3cla & 213 & 1003030001001 & 001 & 8137 \\
2ayh & 214 & 1002019001002 & 002 & 8080 \\
1gpc & 217 & 1002026004007 & 003 & 7920 \\
1akz & 223 & 1003011001001 & 001 & 7607 \\
1dad & 224 & 1003025001005 & 001 & 7555 \\
1aol & 227 & 1002015001001 & 001 & 7404 \\
1cby & 227 & 1004058001001 & 001 & 7404 \\
1lbd & 238 & 1001087001001 & 001 & 6874 \\
2baa & 243 & 1004002001001 & 001 & 6638 \\
1mrj & 247 & 1004094001001 & 001 & 6453 \\
3fib & 248 & 1004098001001 & 001 & 6407 \\
1plq & 258 & 1004076001002 & 001 & 5966 \\
2cba & 258 & 1002050001001 & 002 & 5966 \\
1arb & 262 & 1002031001001 & 001 & 5796 \\
1ako & 268 & 1004086001001 & 001 & 5549 \\
2dri & 271 & 1003072001001 & 001 & 5428 \\
1tml & 286 & 1003002001001 & 001 & 4842 \\
1han & 287 & 1004020001003 & 002 & 4803 \\
1nar & 289 & 1003001001005 & 002 & 4728 \\
1amp & 290 & 1003052003004 & 001 & 4691 \\
1ctt & 294 & 1003075001001 & 001 & 4550 \\
2ctc & 307 & 1003052003001 & 001 & 4107 \\
1ede & 310 & 1003050001003 & 001 & 4007 \\
1pgs & 311 & 1002011001001 & 001 & 3974 \\
1ads & 315 & 1003001005001 & 002 & 3849 \\
1hyt & 316 & 1001053001001 & 002 & 3818 \\
1tca & 317 & 1003050001007 & 001 & 3788 \\
1pot & 321 & 1003073001001 & 011 & 3675 \\
1axn & 323 & 1001051001001 & 001 & 3620 \\
1dxy & 329 & 1003013009001 & 002 & 3463 \\
1nif & 333 & 1002005001003 & 001 & 3362 \\
1rpa & 341 & 1003043001002 & 001 & 3169 \\
1uby & 348 & 1001091001001 & 001 & 3007 \\
1idk & 359 & 1002056001002 & 001 & 2764 \\
1eur & 360 & 1002045001001 & 004 & 2742 \\
1cem & 363 & 1001073001002 & 001 & 2681 \\
1pud & 372 & 1003001017001 & 001 & 2509 \\
1kaz & 377 & 1003041001001 & 001 & 2418 \\
1edg & 380 & 1003001001003 & 002 & 2366 \\
1php & 394 & 1003066001001 & 003 & 2141 \\
1phc & 405 & 1001075001001 & 001 & 1975 \\
1uae & 417 & 1004035002001 & 001 & 1806 \\
1gnd & 430 & 1003004001003 & 001 & 1636 \\
1csh & 433 & 1001074001001 & 001 & 1599 \\
1pmi & 440 & 1002058002001 & 001 & 1521 \\
1gcb & 452 & 1004003001001 & 008 & 1400 \\
2bnh & 456 & 1003007001001 & 001 & 1363 \\
3grs & 461 & 1003004001004 & 001 & 1322 \\
1gai & 471 & 1001073001001 & 001 & 1251 \\
1lam & 484 & 1003036001001 & 001 & 1172 \\
1vnc & 576 & 1001080001001 & 001 & 711 \\
1ciy & 577 & 1002013001002 & 002 & 706 \\
1amj & 753 & 1003005002001 & 002 & 177 \\
1gpb & 823 & 1003068001002 & 001 & 36 \\
1qba\footnote{The longest protein in the set, 1qba, was
used only as a structural template} & 858 & 1002001001005 & 002 & 0 \\
\hline
\caption{List of ``training proteins'' used to extract interaction
potentials.\label{training.table}}
\end{longtable}

\onecolumn 
 
\twocolumn

\end{document}